\documentclass[twocolumn,groupedaddress,amsmath,amssymb,prl,longbibliography]{revtex4-1}

\usepackage{amssymb,amsmath}
\usepackage{cmap}
\usepackage{graphicx}
\usepackage{bm}
\usepackage{color}
\usepackage{blindtext}
\usepackage{hyperref}
\usepackage{listings}
\usepackage{booktabs}
\usepackage{multirow}
\usepackage{amsmath}
\usepackage{mathrsfs}
\usepackage{braket}

\usepackage{dsfont}
\usepackage{tabularx}

\begin{document}

\title{Spatiotemporal Vortex Pulses: Angular Momenta and Spin-Orbit Interaction}

\author{Konstantin Y. Bliokh}
\affiliation{Theoretical Quantum Physics Laboratory, RIKEN Cluster for Pioneering Research, Wako-shi, Saitama 351-0198, Japan}


\begin{abstract}
Recently, spatiotemporal optical vortex pulses carrying a purely transverse intrinsic orbital angular momentum were generated experimentally [{\it Optica} {\bf 6}, 1547 (2019); {\it Nat. Photon.} {\bf 14}, 350 (2020)]. However, an accurate theoretical analysis of such states and their angular-momentum properties remains elusive. Here we provide such analysis, including scalar and vector spatiotemporal Bessel-type solutions as well as description of their propagational, polarization, and angular-momentum properties. Most importantly, we calculate both local densities and integral values of the spin and orbital angular momenta, and predict observable spin-orbit interaction phenomena related to the coupling between the transverse spin and orbital angular momentum. Our analysis is readily extended to spatiotemporal vortex pulses of other natures (e.g., acoustic).
\end{abstract}


\maketitle

{\it Introduction.---}
Vortex beams carrying intrinsic orbital angular momentum (OAM) are widely studied and exploited in modern optics \cite{Allen1992, Allen_book, Bekshaev_book, Andrews_book, Molina2007, Franke2008, Bliokh2015PR}, acoustics \cite{Marston1999, Lekner2006, Volke2008, Demore2012, Anhauser2012, Marzo2018},
and electron microscopy \cite{Bliokh2007, Uchida2010, Verbeeck2010, McMorran2011, Bliokh2017, Lloyd2017, Larocque2018}.
They have found numerous applications in a variety of classical and quantum systems.
Such beams are monochromatic, and their intrinsic OAM is produced by a screw phase dislocation (vortex) aligned with the beam axis \cite{Nye1974, Berry1981}. Thus, this OAM is {\it longitudinal}, i.e., aligned with the mean momentum of the beam. 

Recently, spatiotemporal analogues of vortex beams, {\it spatiotemporal vortex pulses (STVPs)}, were predicted theoretically \cite{Sukhorukov2005, Dror2011, Bliokh2012} and generated experimentally in optics \cite{Jhajj2016,Hancock2019,Chong2020,Hancock2020,Wan2021}. Such states are essentially polychromatic, and they carry intrinsic OAM {\it transverse} (or, generally, tilted) with respect to the propagation direction of the pulse. This OAM is produced by an edge (or mixed edge-screw) phase dislocation \cite{Nye1974, Berry1981}. 
It is anticipated that the STVPs and transverse intrinsic OAM can considerably extend functionality and applications of wave vortices. 
{In particular, while monochromatic vortex beams are essentially 3D objects, STVPs can exist in 2D spatial geometry. In addition, they can produce temporal analogues of spatial phenomena known for monochromatic vortex beams (e.g., time delays instead of beam shifts).} 

Despite the very recent progress in the generation of STVPs \cite{Hancock2019,Chong2020,Hancock2020,Wan2021}, they still lack an accurate theoretical description, including consistent analysis of their angular-momentum and polarization properties. Indeed, despite numerous mentionings of the OAM in Refs. \cite{Hancock2019,Chong2020,Hancock2020,Wan2021}, its value has not been obtained there. Instead, the topological number of the phase dislocation, $\ell$, was used, which generically does not coincide with the normalized OAM value \cite{Allen_book,Berry1998}. Furthermore, accurate calculations of the OAM are impossible without a full vector description and separation of the spin and orbital parts of the total angular momentum \cite{Bliokh2015PR, Bliokh2010, Barnett2010, BB2011}. 

In this paper, we fill this gap by constructing simple Bessel-type solutions for STVPs. We describe their propagational dynamics including `temporal diffraction' \cite{Bliokh2012,Hancock2019}, examine polarization properties, and calculate both local densities and integral values of the OAM and spin angular momentum (SAM) of such pulses. We show that the in-plane linear polarization inevitably produces a longitudinal field component and a nonzero transverse SAM density. This produces the {\it spin-orbit interaction} effects, known for monochromatic beams \cite{Bliokh2015NP,Bliokh2010}, such as observable polarization-dependent intensity distributions of STVPs. Importantly, an integral value of the transverse SAM vanishes, while the integral OAM is quantized as $\hbar\ell$ per photon only for circularly-symmetric pulses with equal width and length. For {\it elliptical} STVPs with different width and length, which were used in experiments  \cite{Hancock2019,Chong2020,Hancock2020,Wan2021}, the OAM value is larger than $\hbar\ell$ per photon.

Thus, our work provides the self-consistent full-vector description of optical STVPs. It also allows straightforward extension to analogous acoustic pulses.

\begin{figure}[!t]
\includegraphics[width=\linewidth]{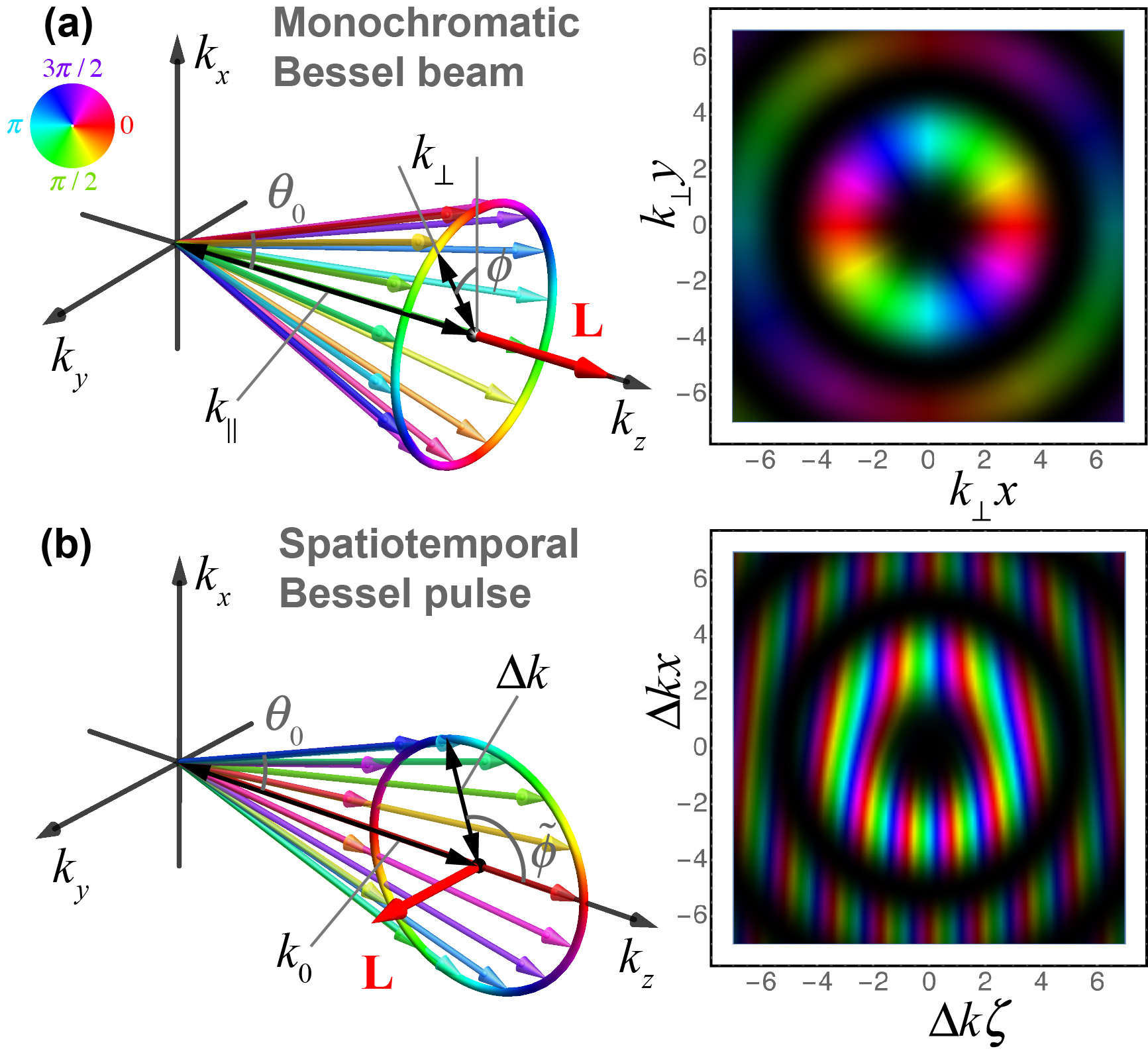}
\caption{The plane-wave spectra (left) and phase-intensity distributions of real-space wavefunctions $\psi({\bf r},t)$ (right) for (a) the monochromatic Bessel beam with $\ell=2$ and (b) spatiotemporal Bessel pulse with $\ell=2$, Eqs.~(\ref{eq1}) and (\ref{eq2}).
In real-space distributions, the brightness is proportional to the intensity $|\psi|^2$, while the color indicates the phase ${\rm Arg}(\psi)$. 
\label{Fig1}}
\end{figure}

{\it Scalar Bessel-type solutions.---}
We first consider scalar waves and simplest analytical vortex-beam solutions: Bessel beams \cite{Durnin1987, Durnin1987_II, McGloin2005, Bouchal1995, Jauregui2005, Bliokh2010}. Monochromatic Bessel beams propagating along the $z$-axis can be constructed as a superposition of plane waves with the same frequency $\omega=\omega_0$, wavevectors ${\bf k}$ distributed within a cone of polar angle $\theta = \theta_0$, and with an azimuthal phase difference $\ell\phi$ ($\phi$ is the azimuthal angle in ${\bf k}$-space) corresponding to a vortex of integer order $\ell$, Fig.~\ref{Fig1}(a). In other words, the wavevectors form a circle 
in the $k_z = k_\parallel$ plane, with the center at $(0,0,k_\parallel)$ and radius $k_\perp$, where $k_\parallel = k_0 \cos\theta_0$, $k_\perp = k_0 \sin\theta_0$, $k_0 = \omega_0/c$, and $c$ is the speed of light. In real space, this superposition results in the scalar wavefunction $\psi({\bf r},t) \propto J_{\ell} (k_\perp r) \exp (i k_\parallel z + i\ell \varphi - i\omega_0 t)$, where $(r,\varphi)$ are the polar coordinates in the $(x,y)$-plane and $J_n$ is the Bessel function of the first kind. The transverse intensity and phase distributions of such Bessel beam are shown in Fig.~\ref{Fig1}(a).

To construct a Bessel-type STVP with a purely transverse intrinsic OAM, we use a superposition of plane waves with wavevectors distributed over a circle in the $k_y=0$ plane with the center at $(0,0,k_0)$ and radius $\Delta k$, Fig.~\ref{Fig1}(b). Using the azimuthal angle $\tilde\phi$ with respect to the center of this circle, we introduce the azimuthal phase difference $\ell\tilde\phi$ and write the real-space wavefunction as a Fourier-type integral:
\begin{equation}
\label{eq1}
\psi({\bf r},t)\! \propto\! \int_0^{2\pi} e^{i [ {k_0 z}+\Delta k \cos\tilde\phi\, z +\Delta k \sin\tilde\phi \, x + {\ell\, \tilde \phi} - \omega ( \tilde\phi ) t ] } \, d \tilde\phi \, ,
\end{equation}
where $\omega ( \tilde\phi ) = c\sqrt{k_0^2 +\Delta k^2 + 2k_0\Delta k \cos \tilde\phi}$. Parameter $\Delta k$ determines the degree of paraxiality and monochromaticity of the Bessel STVP. The maximum polar angle of the wavevectors in its spectrum is $\sin\theta_0 = \Delta k/k_0$, Fig.~\ref{Fig1}(b).   For $\Delta k \ll k_0$, $\theta_0 \ll 1$, the pulse can be considered as near-paraxial and quasi-monochromatic. Below we will use this approximation keeping terms linear in $\theta_0$, which describe some post-paraxial phenomena. 

In this approximation, $\omega ( \tilde\phi ) \simeq c(k_0+ \Delta k \cos \tilde\phi)$, and the integral (\ref{eq1}) results in the analytical Bessel-pulse solution:
\begin{equation}
\label{eq2}
\psi({\bf r},t)\! \propto J_{\ell} (\tilde\rho) \exp (i k_0 \zeta + i\ell \tilde\varphi) \, .
\end{equation}
Here, $\zeta = z - ct = \tilde r \cos\tilde\varphi$, $\tilde\rho = \Delta k \tilde{r}$, and $(\tilde r, \tilde\varphi)$ are the polar coordinates in the $(\zeta,x)$ plane. The intensity $I = |\psi|^2$ and phase ${\rm Arg}(\psi)$ distributions for the Bessel STVP (\ref{eq2}) are shown in Fig.~\ref{Fig1}(b). It has typical Bessel-beam intensity profile $I \propto |J_{\ell}(\tilde\rho)|^2$ in the $(\zeta,x)$ plane, contains an edge phase dislocation of order $\ell$, and propagates with the speed of light along the $z$-axis.

\begin{figure}[!t]
\includegraphics[width=\linewidth]{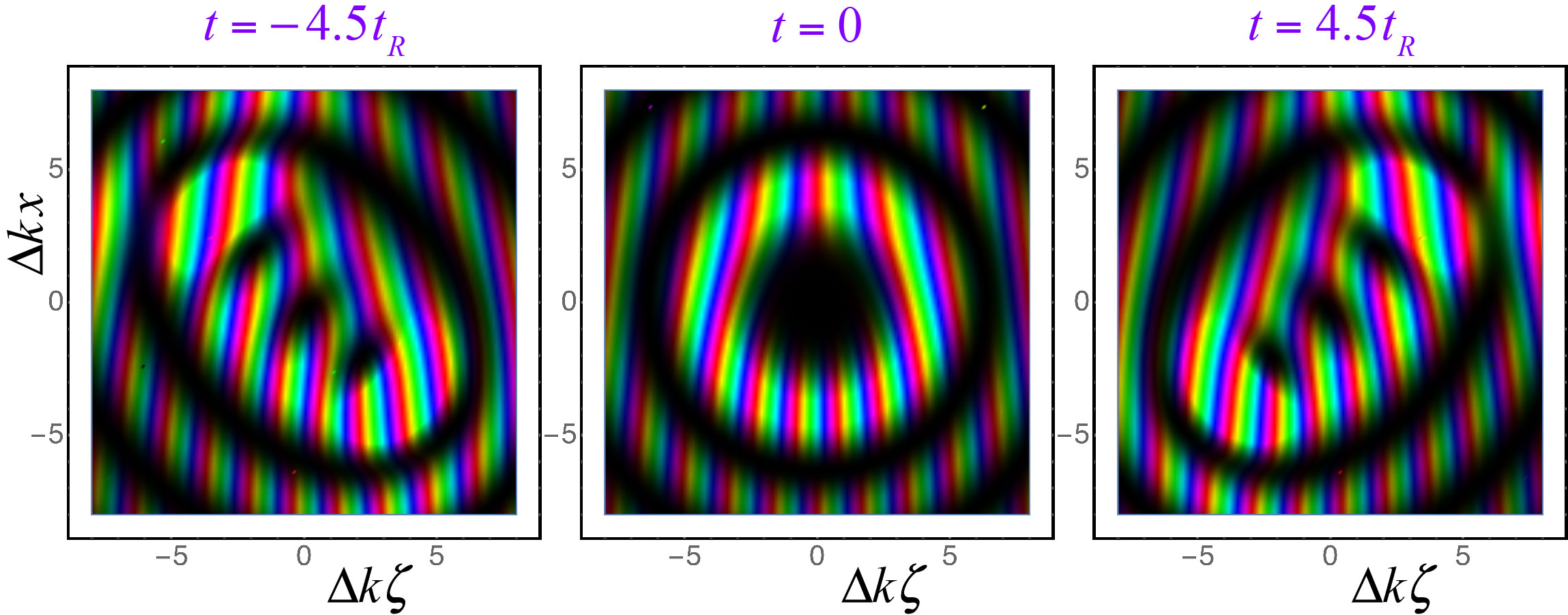}
\caption{Temporal diffraction of the spatiotemporal Bessel pulse (\ref{eq1}) with $\ell=3$ and $\Delta k/k_0 = 0.3$. The characteristic time scale $c t_R = k_0/\Delta k^2$ is analogous to the Rayleigh range of spatially diffracting beams. 
\label{Fig2}}
\end{figure}

Importantly, the nondiffracting solution (\ref{eq2}) is a result of linear expansion  of $\omega ( \tilde\phi )$ with respect to $\Delta k$. The exact solution (\ref{eq1}) evolves in time as shown in Fig.~\ref{Fig2}. Namely, the $\ell$th order phase dislocation in the pulse center splits into a raw of $|\ell |$ first-order dislocations oriented diagonally in the $(\zeta,x)$ plane. This `{\it temporal diffraction}' was predicted in Ref.~\cite{Bliokh2012} and observed in \cite{Hancock2019}. Akin to the Rayleigh range characterizing spatial diffraction, a typical scale of the temporal diffraction is given by the `{\it temporal Rayleigh range}' $c t_R = k_0/\Delta k^2$, Fig.~\ref{Fig2}.
Notably, nondiffracting Bessel-like STVPs can be constructed using the wavevectors distributed along an ellipse in ${\bf k}$ space which could be Lorentz-transformed to a monochromatic circle \cite{Bliokh2012}. 
{In our case, the circular spectrum in Fig.~\ref{Fig1}(b)  should be entended along the $k_z$-axis to become an ellipse with the ratio of semiaxes $\gamma = \sqrt{1+(k_0 /\Delta k)^2}$.}

\begin{figure}[t!]
\includegraphics[width=0.9\linewidth]{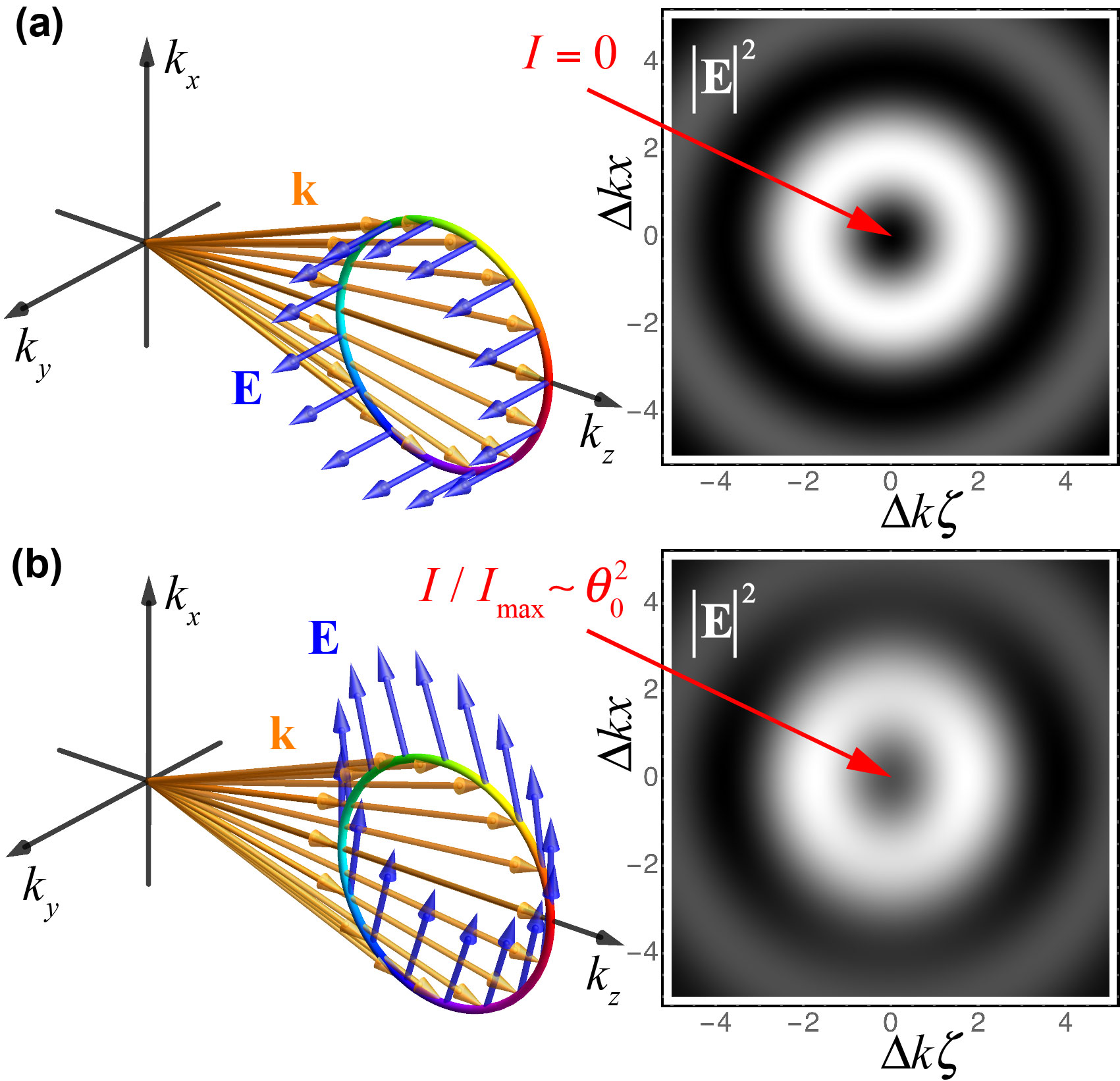}
\caption{Electric fields of plane waves in the spectra of Bessel-type vector STVPs (left) and the corresponding real-space intensity distributions $|{\bf E}({\bf r},t)|^2$ (right) for (a) the out-of-plane polarization and (b) the in-plane polarization. The parameters are: $\ell =1$ and $\Delta k =0.7$ for better visibility of nonzero intensity (\ref{eq5}) in the center of the pulse in (b). 
\label{Fig3}}
\end{figure}

{\it Vector solutions and spin-orbit effects.---}
We now examine vector Bessel STVPs. For simplicity, we consider the electric field ${\bf E}$ of transverse electromagnetic waves; similar arguments could be applied to the magnetic field and other types of vector waves. Due to the transversality condition, the electric field of each plane wave in the pulse spectrum must be orthogonal to its wavevector ${\bf k}$. This determines two basic polarizations in the problem: (i) out-of-plane, ${\bf E}$ is directed along the $y$-axis, Fig.~\ref{Fig3}(a), and (ii) in-plane, ${\bf E}$ lies in the $(z,x)$ plane, Fig.~\ref{Fig3}(b).

For the out-of-plane polarization, the field has only one component, and the problem reduces to the scalar case: ${E_y}({\bf r},t) \propto \psi({\bf r},t)$.

For the in-plane polarization, the situation is less trivial. Each plane wave in the pulse spectrum has two electric-field components, $E_x$ and $E_z$, Fig~\ref{Fig3}(b). The amplitudes and phases of these components depend on the wavevector ${\bf k}$, which signals the {\it spin-orbit interaction} \cite{Bliokh2015NP,Bliokh2010}.
To describe the spin-orbit effects in the linear approximation in $\Delta k$, we write the electric field components for each plane wave as $E_x = E\cos\theta \simeq E$ and $E_z = - E \sin\theta \simeq - E (\Delta k/k_0)\, {\sin\tilde\phi}$, where $\theta \leq \theta_0$ is the polar angle of a given wavevector. Since for the integrand of Eq.~(\ref{eq1}), $i\Delta k\, {\sin \tilde\phi} = \partial/\partial x$, one can write the real-space transverse ($x$) and longitudinal ($z$) field components as
\begin{equation}
\label{eq3}
E_x ({\bf r},t) \simeq \psi ({\bf r},t), \quad
E_z ({\bf r},t) \simeq \frac{i}{k_0}\frac{\partial \psi ({\bf r},t)}{\partial x}
 \, .
\end{equation}
%

Substituting Eq.~(\ref{eq2}) into Eq.~(\ref{eq3}), we obtain the longitudinal field 
\begin{equation}
\label{eq4}
E_z \!\propto\! \frac{i\Delta k}{2} e^{i k_0 \zeta} \!\left[e^{i(\ell-1)\tilde\varphi} J_{\ell-1}(\tilde{\rho}) + e^{i(\ell+1)\tilde\varphi} J_{\ell+1}(\tilde{\rho})
\right].
\end{equation}
From Eqs.~(\ref{eq2})--(\ref{eq4}), the total field intensity, $I = |E_x|^2 + |E_z|^2$, is given by
\begin{equation}
\label{eq5}
I \propto J_{\ell}^2 + \frac{\Delta k^2}{4k_0^2}
\!\left[J_{\ell-1}^2 + J_{\ell+1}^2
+ 2\cos (2\tilde\varphi) J_{\ell-1} J_{\ell+1}
\right].
\end{equation}
Here and hereafter, for brevity, we omit the Bessel-functions argument $\tilde{\rho}$.

The intensity distribution of the Bessel STVP, Eq.~(\ref{eq5}), resembles intensity distributions of vector Bessel beams in optics \cite{Bliokh2010}, acoustics \cite{Bliokh2019_II}, and quantum mechanics \cite{Bliokh2011PRL}. In particular, the presence of the Bessel functions of orders $\ell \pm 1$ is a typical signature of the spin-orbit interaction. The easiest-to-observe spin-orbit effect is a nonzero intensity $\sim \theta_0^2$ in the center of the in-plane-polarized STVPs with $|\ell|=1$, Fig.~\ref{Fig3}. For monochromatic vortex beams, this phenomenon has been observed in optical experiments \cite{Bokor2005,Gorodetski2008,Bliokh2015NP}.
The main difference is that in the case of monochromatic vortex beams this phenomenon occurs for {\it circular} polarizations, corresponding to the longitudinal SAM of the beam, while for STVPs it takes place for the in-plane {\it linear} polarization. As we show below, this polarization generates a nonzero transverse SAM density directed along the $y$-axis, i.e., along with the intrinsic OAM of the pulse. 

{\it Spin and orbital angular momenta.---}
Analysis of the spin and orbital angular momenta of polychromatic STVPs is a challenging problem because most theoretical methods are developed for monochromatic fields. Indeed, standard formulas for the SAM and OAM densities imply averaging over the cycle of periodic temporal oscillations \cite{Bliokh2015PR}, and they become ill-defined in generic polychromatic fields \cite{BB2011}. Nonetheless, here we can employ the quasi-monochromaticity of pulses with $\Delta k \ll k_0$ and separate fast temporal oscillations with the central frequency $\omega_0 = k_0 c$ and slow temporal evolution with the inverse temporal scale $\Delta k\, c$. This results in the SAM and OAM densities given by the standard monochromatic formulas involving canonical SAM and OAM operators and time-dependent `wavefunction' ${\bf E}({\bf r},t) e^{i\omega_0 t}/\sqrt{\omega_0}$ \cite{Bliokh2015PR}:
\begin{equation}
\label{eq6}
S_y = \omega_0^{-1} {\rm Im} \left( {\bf E}^* \times {\bf E} \right)_y, ~~
L_y = \omega_0^{-1} {\rm Im}\! \left( {\bf E}^*\cdot \frac{\partial}{\partial \tilde\varphi} {\bf E} \right).
\end{equation}
Although the angle $\tilde\varphi$ in the $(\zeta,x)$ plane involves time, in Eq.~(\ref{eq6}) we used the fact that at $t=0$ it becomes the desired azimuthal angle in the $(z,x)$ plane, whereas in the diffractionless approximation the SAM and OAM distributions are invariantly translated together with the pulse along the $z$-axis.

\begin{figure}[t!]
\includegraphics[width=\linewidth]{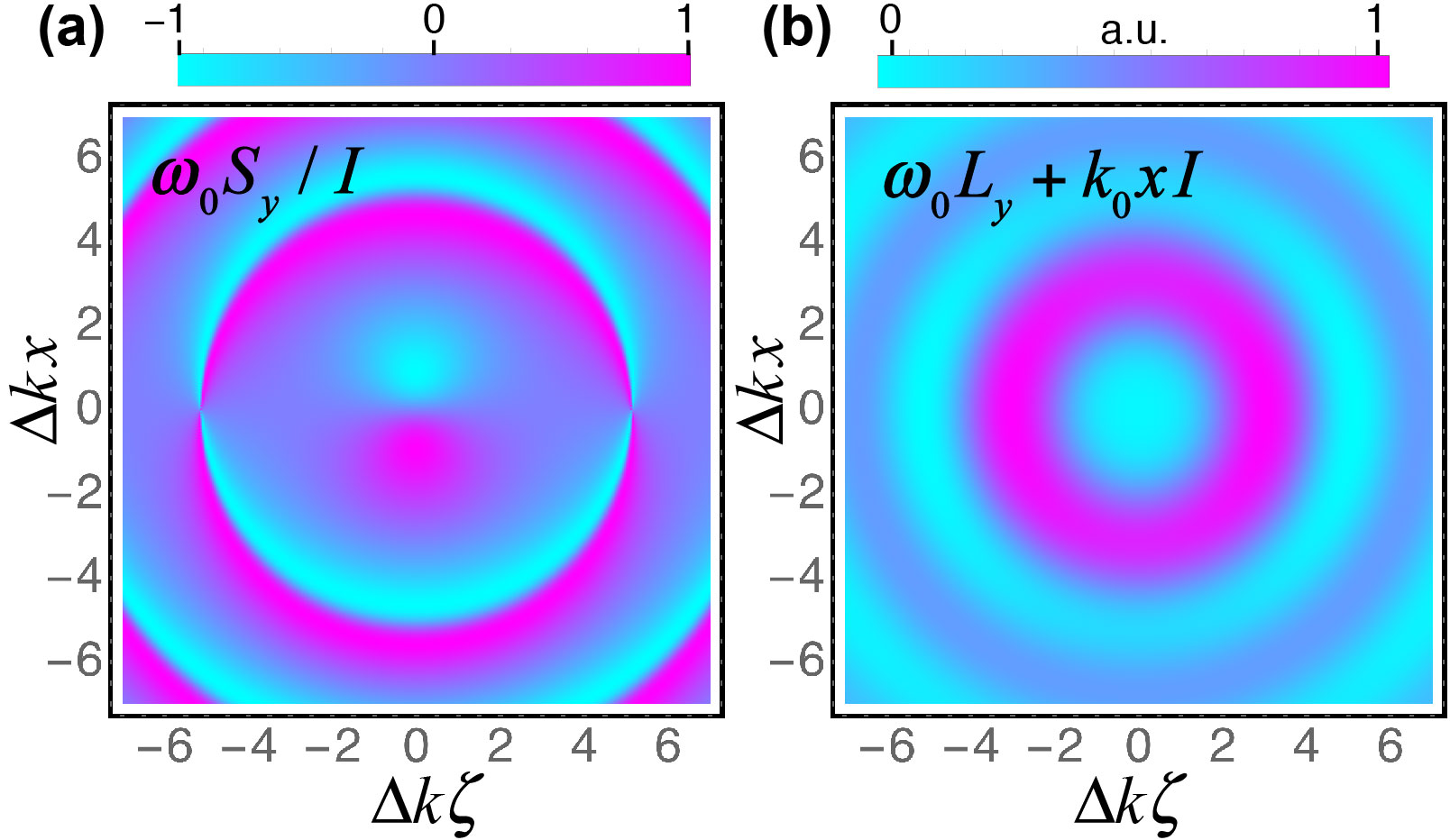}
\caption{Spatial distributions of (a) the normalized transverse spin angular momentum density $\omega_0 S_y/I$ and (b) the orbital angular momentum density (with the subtracted propagational term) $\omega_0 L_y + k_0xI$ in the in-plane-polarized Bessel STVP,	 Eqs.~(\ref{eq8}) and (\ref{eq5}). The parameters are: $\ell=2$ and $\Delta k/k_0 =0.5$. 
\label{Fig4}}
\end{figure}

For the out-of-plane polarization (the scalar case), Eqs.~(\ref{eq6}) yield 
\begin{equation}
\label{eq7}
S_y = 0\,, \quad
L_y = \omega_0^{-1}(\ell - k_0 x)\, I\,.
\end{equation}
where $I\propto J_{|\ell|}^2(\tilde\rho)$. The OAM density (\ref{eq7}) contains the standard vortex-related term proportional to $\ell$ as well as the $k_0 x$-dependent term caused by the $z$-propagation of the pulse. After integration of the OAM density $L_y$ over the $(z,x)$ plane, the $k_0 x$-term vanishes, while the vortex-induced term yields the integral value of $\hbar\ell$ per photon \cite{Allen1992, Allen_book, Bekshaev_book, Andrews_book, Molina2007, Franke2008, Bliokh2015PR}: $\omega_0 \langle L_y \rangle/\langle I \rangle = \ell$. Here $\langle ... \rangle = \int ...\,dz\, dx = \int ...\,d\zeta\, dx$ and we imply quasi-monochromatic quantization of photon's energy, $\hbar \omega_0$.

For the in-plane polarized pulse (\ref{eq3})--(\ref{eq5}), Eqs.~(\ref{eq6}) yield
\begin{align}
\label{eq8}
S_y \propto \omega_0^{-1} \frac{\Delta k^2 x}{2\ell\, k_0} \left(J_{\ell+1}^2 -J_{\ell-1}^2\right), 
\nonumber \\
L_y = \omega_0^{-1}(\ell - k_0 x)\, I + \frac{\ell}{2k_0 x} S_y\,.
\end{align}
These distributions are depicted in Fig.~\ref{Fig4}. The nonzero spin density $S_y$ is a manifestation of the {\it transverse spin} phenomenon, which recently attracted great attention \cite{Bliokh2015PR, Bliokh2015NP, Aiello2015, Lodahl2017, Eismann2021}. Here, the transverse spin arises from the interference of plane waves with different directions, phases, and in-plane linear polarizations, Fig.~\ref{Fig3}(b). The normalized SAM density $\omega_0 S_y/I$ reaches the minimum and maximum values of $-1$ and $1$, Fig.~\ref{Fig4}(a), i.e., the polarization becomes left-hand and right-hand circular in these zones. As is typical for the transverse spin of free-space propagating waves \cite{Bliokh2015PR, Aiello2015, Eismann2021}, its integral value vanishes: $\langle S_y \rangle = 0$.

The OAM density (\ref{eq8}) has a form similar to Eq.~(\ref{eq7}) with an additional spin-related term; this is also a singnature of the spin-orbit interaction \cite{Bliokh2015NP,Bliokh2010}. Nonetheless, in contrast to the analogous spin-dependent OAM parts in monochroimatic beams, the integral value of this term vanishes. (This follows from the relation $\ell \int_0^\infty \left( J_{\ell+1}^2 - J_{\ell-1}^2 \right) \tilde\rho\, d \tilde\rho \propto \ell\int_0^\infty  J_{\ell}d J_{\ell}/d \tilde\rho\, d \tilde\rho = 0$.) Thus, akin to the scalar case, the integral OAM value corresponds to $\hbar\ell$ per photon.  

We conclude that in spite of local spin-orbit interaction effects, the integral SAM and OAM of STVPs are rather robust:
\begin{equation}
\label{eq9}
\langle S_y \rangle = 0\,, \qquad
\frac{\omega_0 \langle L_y \rangle}{\langle I \rangle} = \ell \,.
\end{equation}
%

\begin{figure}[t!]
\includegraphics[width=\linewidth]{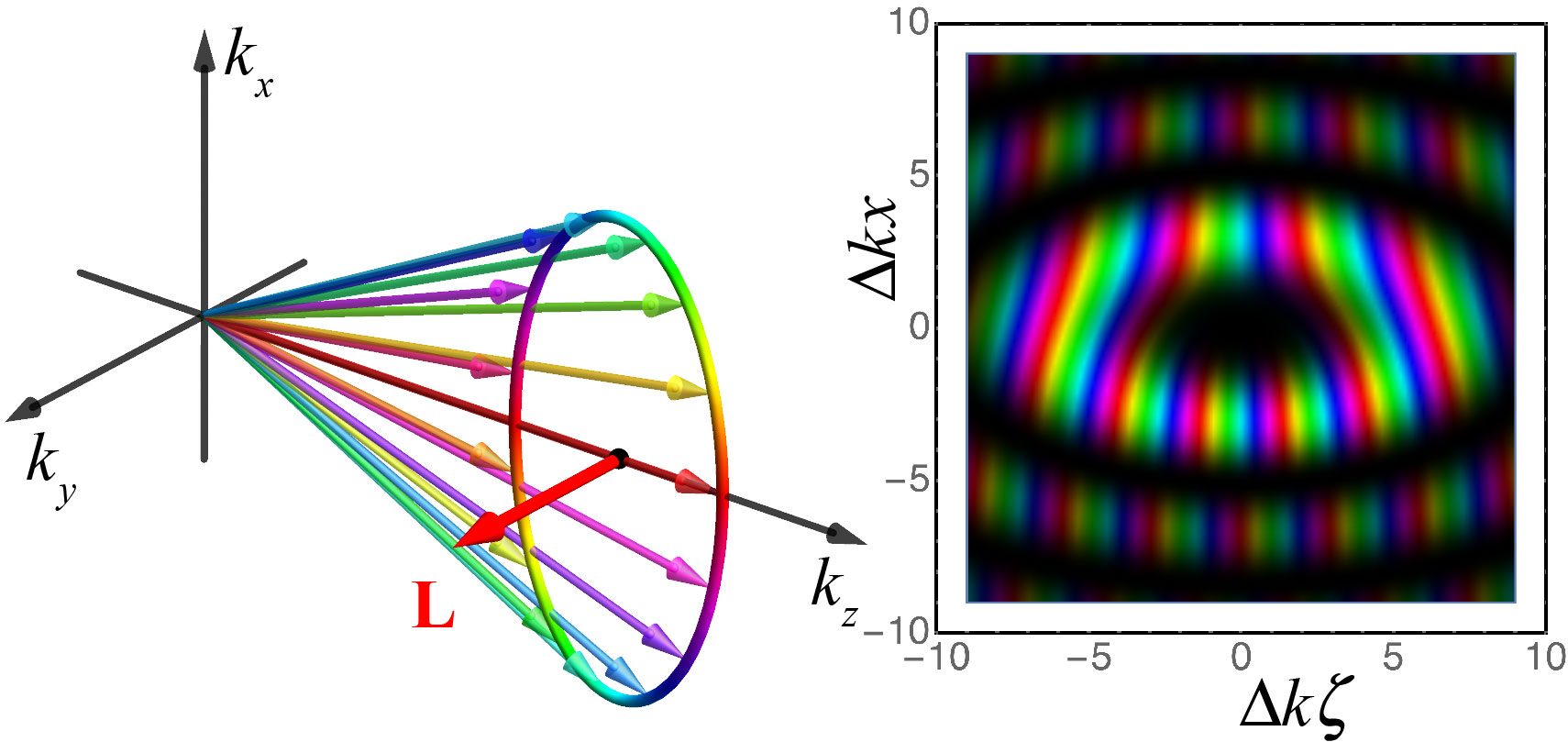}
\caption{The plane-wave spectrum (left) and the phase-intensity distribution of the real-space wavefunction $\psi({\bf r},t)$ (right) for an elliptical Bessel STVP with $\ell=2$ and ratio of principal axes $\gamma = 2$. The normalized integral OAM of this pulse is $\omega_0 \langle L_y \rangle/\langle I \rangle = 2.5$, Eq.~(\ref{eq10}).
\label{Fig5}}
\end{figure}

Importantly, the above calculations are made for STVPs with {\it circularly} symmetric intensity profiles in the $(\zeta,x)$ plane. 
However, in most cases these profiles are {\it elliptical} with some ratio of principal axes $\gamma$ in the $(\zeta,x)$ plane, as shown in Fig.~\ref{Fig5}. Such an elliptical STVP is described by the substitution $x \to \gamma x$ in the scalar wavefucntion (\ref{eq1}) and (\ref{eq2}) or, equivalently, by the substitution $k_x \to \gamma^{-1}k_x$ in the beam spectrum. For the field of the form $\propto \exp(i\ell \tilde\varphi)$ and the OAM operator $\hat{L}_y = -i \partial/\partial \tilde\varphi = i\left(x \partial / \partial \zeta - \zeta \partial / \partial x \right)$, this results in an additional factor in the intrinsic OAM value \cite{Bliokh2012,Bliokh2012PRL}:
\begin{equation}
\label{eq10}
\frac{\omega_0 \langle L_y \rangle}{\langle I \rangle} = \frac{\gamma+\gamma^{-1}}{2}\, \ell \,.
\end{equation}
%
This factor is significant: for example, the experiments \cite{Hancock2019,Chong2020} generated STVPs with $\gamma \simeq 2.5$--$3$, which yields $(\gamma+\gamma^{-1})/2 \simeq 1.5$--$1.7$. Figure~\ref{Fig5} shows an example of the STVPs with $\ell=2$, $\gamma=2$, and $\omega_0 \langle L_y \rangle/\langle I \rangle = 2.5$.

{\it Conclusions.---}
We have examined spatiotemporal vortex pulses with purely transverse intrinsic orbital angular momentum. We provided analytical Bessel-type solutions, both scalar and vector, and described their propagation, polarization, and angular-momentum properties. Most importantly, we provided accurate calculations of the spin and orbital angular momenta of STVPs and described observable spin-orbit interaction phenomena.
Notably, the polarization and spin-orbit effects manifest themselves locally via observable intensity and spin-density distributions, while the integral values of the spin and orbital angular momenta of STVPs are rather robust. At the same time, the integral OAM value is significantly affected by the elliptical shape of STVPs with different width and length, which are typically generated in experiments \cite{Hancock2019,Chong2020}. 

The results of our work provide a theoretical platform for investigations of novel spatiotemporal vortex states and call for experimental measurements of the predicted polarization and angular-momentum phenomena. 
{We considered Bessel-type pulses solely for the simplicity of their theoretical description. The results can be straightforwardly generalized to the Laguerre-Gaussian-type spectra \cite{Allen1992,Allen_book,Bekshaev_book,Andrews_book}, more relevant to typical experimental situations. 
Furthermore, a variety of phenomena, well studied for monochromatic vortex beams, can now be investigated for STVPs: e.g., fractional OAM \cite{Berry2004,Leach2004_II}, shifts at planar interfaces \cite{Bliokh2009_II,Merano2010}, etc.}

Importantly, our results are applicable to waves of different natures. For example, STVPs can be generated in sound waves in fluids or gases. In doing so, one can use the scalar approach (\ref{eq1})--(\ref{eq2}) for the pressure wavefield $P({\bf r},t)$ or the vector approach similar to Eqs.~(\ref{eq3})--(\ref{eq5}) for the velocity wavefield ${\bf V}({\bf r},t)$. Since sound waves are {\it longitudinal}, i.e., ${\bf V} \parallel {\bf k}$ for each plane wave in the spectrum, the velocity field has the $z$ and $x$ components, {$V_z ({\bf r},t) \simeq \psi ({\bf r},t)$ and $V_x ({\bf r},t) \simeq - i k_0^{-1} \partial \psi ({\bf r},t)/\partial x$}, generating the transverse spin and related spin-orbit phenomena \cite{Shi2019,Bliokh2019_II}. 



\bibliography{References}

\end{document}